\documentclass[reprint,prl]{revtex4-1}
\usepackage{amsmath,amssymb,bm}
\usepackage{graphicx}

\begin{document}
 \title{Mirror winding number and helical edge modes in honeycomb lattice with hopping-energy texture}

 \date{\today}
 \author{Toshikaze Kariyado}\email{kariyado.toshikaze@nims.go.jp}
 \author{Xiao Hu}\email{HU.Xiao@nims.go.jp}
 \affiliation{International Center for Materials Nanoarchitectonics (WPI-MANA),
 National Institute for Materials Science, Tsukuba 305-0044, Japan}
 % 73.22.Pr Graphene electronic structure
 % 73.43.-f quantum Hall effect
 % 03.65.Vf, Topological phases (quantum mechanics)

  \begin{abstract}
   We illustrate possible topological phases in honeycomb lattice with
   textures in electron hopping energy between
   \textit{nearest-neighboring} sites and show that they are
   characterized by the mirror winding number intimately related to the
   chiral (or sublattice) symmetry. Analytic wave functions of
   zero-energy edge modes in ribbon geometry are provided, which
   are classified into even and odd sectors with respect to the mirror
   operation with the mirror plane perpendicular to the edge,
   and evolve into the topological helical edge states at finite
   momenta. Intriguingly our results demonstrate that in order to
   achieve the topological phase one can decorate the edge in a way
   adaptive to the bulk hopping texture. This paves a new way to
   tailoring graphene in the topological point of view.
  \end{abstract}

 \maketitle
 \noindent\textit{Introduction---}
 Currently topological states of matter are attracting much
 attention \cite{RevModPhys.82.3045,RevModPhys.83.1057}. One aspect of
 the study is to elaborate classification of material phases using
 topological invariants. The other aspect is application oriented and
 topologically protected surface/edge modes are focused. While topology
 is a notion that is not limited by symmetry, symmetries enrich
 topological phase diagrams
 \cite{PhysRevLett.95.146802,PhysRevLett.95.226801,DOI:10.1063/1.3149495,1367-2630-12-6-065010,PhysRevLett.106.106802,PhysRevB.88.125129,PhysRevB.90.165114}. The
 quantum spin Hall effect in electron systems characteristic of
 time-reversal symmetry is the first example of topological state with
 $\mathbb{Z}_2$ topological index. However, if we attempt to design $\mathbb{Z}_2$ topological insulators in a system without spin, it
 is not straightforward to mimic a time-reversal operator squaring to
 $-1$, which plays a key role in the topological
 classification \cite{1367-2630-12-6-065010}. A possible way to
 resolve this difficulty is to relax symmetry to an approximate
 one \cite{PhysRevLett.114.223901,Mousavi:2015aa,Wu:2016aa},
 and it is found that sometimes a symmetry holding in a limited region in
 the Brillouin zone is sufficient to support desired topological phase.

Of various possibilities, modulating honeycomb lattice is expected to be
fruitful, and pedagogical as well as learned from the lessons so far
\cite{PhysRevLett.61.2015,PhysRevLett.95.146802,PhysRevLett.95.226801,Guinea:2010aa,PhysRevLett.109.055502,1367-2630-15-6-063031}. The
idea was first demonstrated in a photonic crystal made of dielectric
cylinders \cite{PhysRevLett.114.223901}, which are aligned in a slightly
distorted honeycomb structure, and then extended to electronic
tight-binding (TB) model in honeycomb lattice with texture in hopping
energy \cite{Wu:2016aa}. When the hopping energy within
hexagons is different from that between hexagons, one can think of molecular orbitals in the hexagonal artificial atoms, and meanwhile
the K and K' points are folded to the $\Gamma$ point in the first
Brillouin zone. These molecular orbitals approximately play a role of
spin, and an operator of pseudo time-reversal symmetry squaring to
$-1$ is achieved. It was revealed that when the inter hexagon hopping
energy is stronger than the intra one, a topological state mimicking the
quantum spin Hall state with time-reversal symmetry can be achieved
without resorting to the intrinsic spin degrees of freedom.

 \begin{figure}[t]
  \begin{center}
   \includegraphics[scale=1.0]{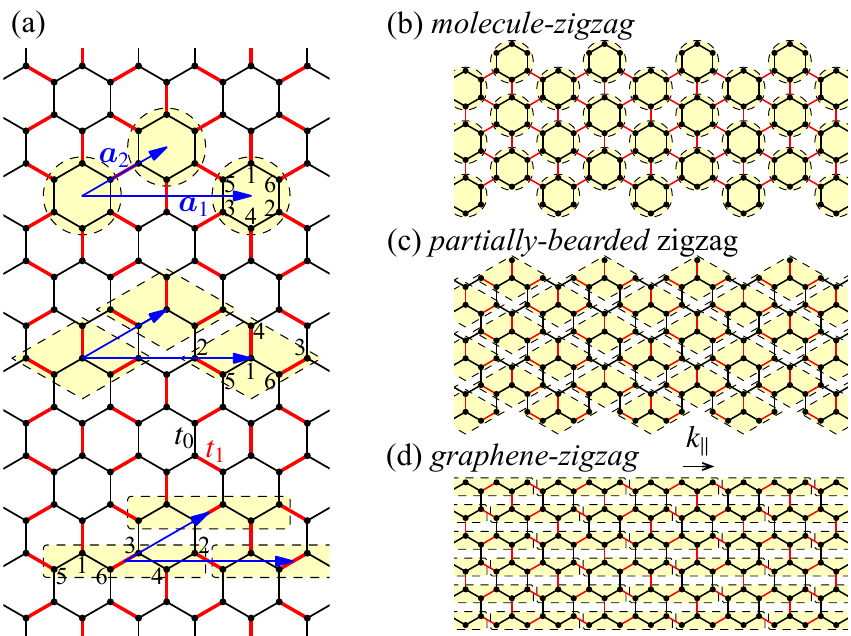}
   \caption{(Color online) (a) TB model in honeycomb lattice with texture in electron hopping energy between nearest-neighboring sites: $t_0$ (thin black bonds) and $t_1$ (thick red bonds) within and among
   hexagons encompassed by the dashed circles, and the two unit vectors
   $\bm{a}_1$ and $\bm{a}_2$. Small numbers in the unit
   cells represent the order for each site appearing in the basis set.
   (b,c,d)
   \textit{molecule-zigzag},
   \textit{partially-bearded} and \textit{graphene-zigzag} edge
   associated with the circular, rhombic and rectangular unit cell
   respectively denoted in (a) by dashed lines.}\label{fig1_lattice}
  \end{center}
 \end{figure}
 In this Letter, we elaborate the discussion on the electronic TB model
 on honeycomb lattice with texture in hopping energy between
 \textit{nearest-neighboring} sites. We reveal that the topological
 state induced
 by the hopping texture can be characterized by the mirror winding
 number, a rigorous topological invariant intimately related to the
 chiral (or sublattice) symmetry. Analytic wave functions are provided
 for edge states at the $\Gamma$ point in ribbon geometry, which are
 classified into even and odd sectors with respect to the mirror
 operation with the mirror plane perpendicular to the edge, and evolve
 into the topological helical edge states at finite momenta, in
 agreement with the results of band calculations. Explicitly we find
 that when the intra-hexagon hopping energy is stronger (weaker) than
 the inter-hexagon one the so-called \textit{molecule-zigzag}
 (\textit{partially-bearded}) edge to honeycomb lattice yields gapless
 helical edge modes (see Fig.~\ref{fig1_lattice}). Our discussions
 provide a new designing guideline for topological edge states, which
 paves a way to rich opportunities for applications.
 
 \medskip
 \noindent\textit{Hamiltonian and chiral symmetry---}
 We consider a TB Hamiltonian in honeycomb lattice 
\begin{equation}
 H=\sum_{\langle{ij}\rangle}t_{ij}c^\dagger_ic_j
\end{equation}
 where $c^\dagger_i$ ($c_i$) is the creation (annihilation) operator at
 site $i$, and
 $\langle{ij}\rangle$ denotes summation over the
\textit{nearest-neighboring} sites, with the hopping integral $t_{ij}$
 modulated spatially (see Ref.~\onlinecite{PhysRevLett.107.066801} for a
 discussion on generic hopping textures).
 To be specific, we concentrate on the case
 that there are two hopping integrals $t_0$ and $t_1$, real and positive
 denoted by the thin black and thick red bonds in
 Fig.~\ref{fig1_lattice} respectively as in the previous work
 \cite{Wu:2016aa}.
 Unit vectors associated with the hopping texture
 are given by $\bm{a}_1$ and $\bm{a}_2$. Because of the bipartite
 nature, the Hamiltonian can be rewritten into the form
\begin{equation}
  H_{\bm{k}}=
   \begin{pmatrix}
    0&Q_{\bm{k}}\\
    Q_{\bm{k}}^\dagger&0
   \end{pmatrix},
   \label{Hamiltonian-k}
\end{equation}
with a basis where the upper (lower) half is for A (B) sublattice and
$Q_{\bm{k}}$ a 3$\times$3 matrix, irrespectively to the choice of unit cell
[see Fig.~\ref{fig1_lattice}(b,c,d)]. 
It is straightforward to confirm that the chiral operator
$\gamma=\mathrm{diag}(1,-1)$ anticommutes with the Hamiltonian
\eqref{Hamiltonian-k}, manifesting the
chiral symmetry (or sublattice symmetry) of the present system, which
provides a powerful tool for one to tackle with the topological
property.

\medskip
\noindent\textit{Mirror winding number---}
Regarding the momentum $\bm{k}$ parallel to the unit vector $\bm{a}_1$ 
as a free parameter, the system can be viewed as an
effective 1D model, to which one can assign the winding number \cite{1367-2630-12-6-065010}. 
At the $\Gamma$ point $k_\parallel=0$, the Hamiltonian
\eqref{Hamiltonian-k} can be reduced to the even and odd sectors by the
mirror symmetry, with the mirror plane perpendicular to $\bm{a}_1$ [see
Fig.~\ref{fig1_lattice}(a)]. As far as the mirror plane is taken to be
compatible with the hopping texture, the reduced Hamiltonian still has
the chiral symmetry since the mirror operation commutes with $\gamma$,
physically meaning that the mirror operation does not mix A and B
sublattices. Then, it is possible to assign winding
numbers for the
even and odd sectors separately
\cite{PhysRevB.88.125129,PhysRevB.90.165114}, which constitutes the
\textit{mirror} winding number $(n_+,n_-)$.

 \begin{table}[tbp]
  \begin{center}
   \caption{Mirror winding number $(n_+,n_-)$ at
   $k_\parallel=0$.  with $\delta\equiv t_1-t_0$ which are defined in Fig.~\ref{fig1_lattice}(a).}\label{table1}
   \begin{tabular}{|c|c|c|}
    \hline
    $(n_+,n_-)$ & $\delta>0$ & $\delta<0$ \\
    \hline
    \textit{molecule-zigzag}& (1,$-1$) & (0,0) \\ \hline
    \textit{partially-bearded}& (0,0) & ($-1$,1) \\ \hline
    \textit{graphene-zigzag}& (1,0)& (0,1)\\ \hline
   \end{tabular}
  \end{center}
 \end{table}

In general, a winding number depends on how the unit cell is taken which
determines the way momentum appearing in the Hamiltonian. Explicitly we
consider three kinds of unit cell and the associated edges, noting that
the edge should not destroy the unit cell for the purpose of discussion
on edge states protected by the bulk topology \cite{PhysRevB.88.245126}. 
The first unit cell is the circular one in Fig.~\ref{fig1_lattice}(a),
and the corresponding edge is given in Fig.~\ref{fig1_lattice}(b)
\cite{doi:10.1021/jp990510j}, which we call \textit{molecule-zigzag} edge
since a hexagonal cluster can be regarded as an artificial molecule
\cite{PhysRevLett.114.223901,Wu:2016aa}. In this case one has
\begin{equation}
 Q_{\bm{k}}=
  \begin{pmatrix}
   t_1X\bar{Y}^2&t_0&t_0\\
   t_0&t_1\bar{X}Y&t_0\\
   t_0&t_0&t_1Y
    \end{pmatrix}
\end{equation}
where $X=e^{i\bm{k}\cdot\bm{a}_1}$ and $Y=e^{i\bm{k}\cdot\bm{a}_2}$. With the mirror symmetry at $k_{\parallel}=0$ , $Q_{\bm{k}}$ is decomposed into
\begin{equation}
 Q^+_{k_\perp}=
  \begin{pmatrix}
   t_1\bar{Y}^2&\sqrt{2}t_0\\
   \sqrt{2}t_0& t_0+t_1Y
  \end{pmatrix},\quad
  Q^-_{k_\perp}=t_1Y-t_0,
\end{equation}
with $+$ and $-$ for the even and odd sector respectively upon the mirror
operation. One has nonzero winding numbers in both even and odd sectors
$(n_+,n_-)=(1,-1)$ for $t_1>t_0$, and zero ones for $t_1<t_0$ (see
appendix for details), which is consistent with the
conclusion in the previous work \cite{Wu:2016aa}.

The second unit cell is the rhombic one in Fig.~\ref{fig1_lattice}(a),
and the associated edge is given in Fig.~\ref{fig1_lattice}(c)
\cite{doi:10.1021/jp990510j}, which we
called \textit{partially-bearded} edge since the top edge is constructed
by adding one extra lattice site every three sites at the zigzag edge
known in the study of graphene. Note that with this unit cell the top
and bottom edges of a ribbon system have different shapes. In this case,
one has
\begin{equation}
Q_{\bm{k}}=
\begin{pmatrix}
 t_1 & t_0 & t_0 \\
 t_0Y & t_1 & t_0X\bar{Y}\\
 t_0\bar{X}Y & t_0\bar{Y} & t_1
\end{pmatrix},
\end{equation}
which is decomposed into the even and odd sectors as
\begin{equation}
 Q^+_{k_\perp}=
  \begin{pmatrix}
   t_1&\sqrt{2}t_0\\
   \sqrt{2}t_0Y&t_1+t_0\bar{Y}
    \end{pmatrix},\quad
    Q^-_{k_\perp}=t_1-t_0\bar{Y}.
\end{equation}
We have $(n_+,n_-)=(-1,1)$ for $t_1<t_0$ and zero winding number for
$t_1>t_0$ (see appendix for details), in contrary to the
first case.

 \begin{figure*}[t]
  \begin{center}
   \includegraphics[scale=1.0]{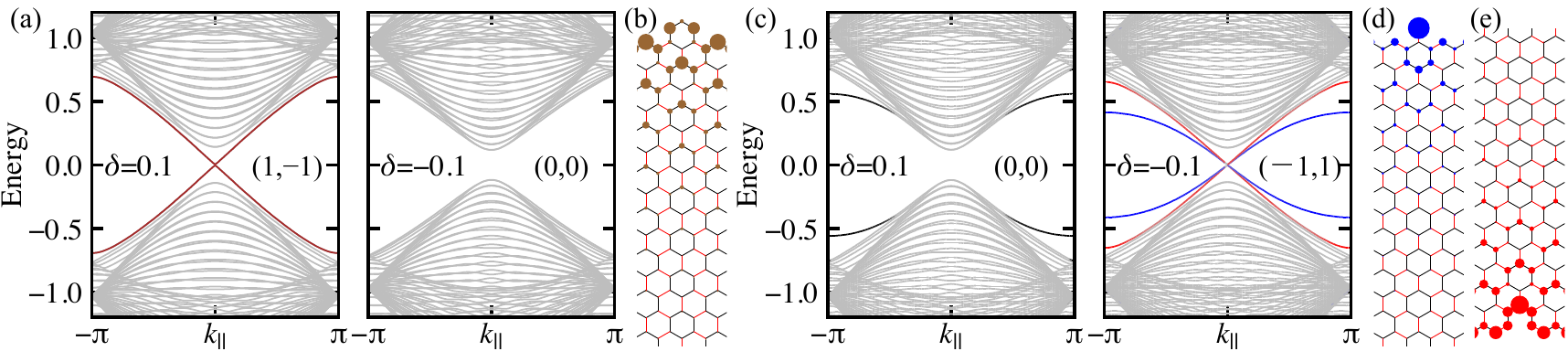}
   \caption{(a,c) Band structures of a
   ribbon system for
   \textit{molecule-zigzag} edge and \textit{partially-bearded} edge,
   with the mirror winding numbers $(n_+,n_-)$ indicated. In (a) the
   helical edge states are highlighted by the brown (dark) dispersions
   for $\delta=0.1$ with double degeneracy, whereas in (c) the blue
   (red) dispersions are for the
   helical edge states localized at the top (bottom) edge,
   with the corresponding real-space distribution of wave
   function $|\psi_i|^2$ displayed in (b,d,e). The
   gray lines in (a) and (c) are for bulk states. The
   black (dark) lines in (c) for $\delta=0.1$ represents edge states
   whose origin is not topology. The ribbon is
   infinitely long along $\bm{a}_1$ with periodic boundary condition and
   contains 40 unit cells along $\bm{a}_2$.}
   \label{fig2_molzig+beard}
  \end{center}
 \end{figure*}

 For the sake of comparison, a third unit cell is also considered as the
 rectangular one in Fig.~\ref{fig1_lattice}(a), corresponding to the
 \textit{graphene-zigzag} edge given in Fig.~\ref{fig1_lattice}(d). It
 is easy to check that $(n_+,n_-)=(1,0)$ for $t_1>t_0$ and
 $(n_+,n_-)=(0,1)$ for $t_1<t_0$  (see appendix materials for
 details). \textit{Armchair} edge is also well known in
 the study of graphene. It is easy to check that one cannot
 assign the mirror winding number for armchair edge, since the mirror
 operation mixes A and B sublattices, which corresponding the absence of
 gapless helical edge states.

 Table~\ref{table1} summarizes the mirror winding number $(n_+,n_-)$ for
 the three cases considered above. For \textit{molecule-zigzag} and
 \textit{partially-bearded} edges, the states for $\delta>0$ and
 $\delta<0$ with $\delta\equiv (t_1-t_0)/t_0$ can be distinguished
 topologically in terms of the mirror winding number, but
indistinguishable by the total winding number $n_{\rm tot}=n_++n_-$; for
 \textit{graphene-zigzag} edge, we have $n_{\text{tot}}=1$ irrespective
 of the sign of $\delta$. While the mirror winding number is only
 defined at $k_\parallel=0$, the total winding number $n_{\rm tot}$ is
 well-defined for any momentum, and is conserved as far as the bulk gap
 is not closed. Because a nonzero winding number specifies a zero-energy
 edge state, one can conclude that there is no
 zero-energy edge state for \textit{molecule-zigzag} and
 \textit{partially-bearded} edges at $k_\parallel\neq{0}$, whereas a
 zero-energy edge state for \textit{graphene-zigzag}
 edge at any momentum, which is naturally connected to the discussion
 for graphene without hopping texture \cite{PhysRevLett.89.077002}.
 To close this section, we notice that mirror winding numbers
 have been discussed recently for several other systems
 \cite{PhysRevLett.111.056403,PhysRevLett.113.046401,1742-6596-603-1-012002}.

 \begin{figure}[tbp]
 \begin{center}
  \includegraphics[scale=1.0]{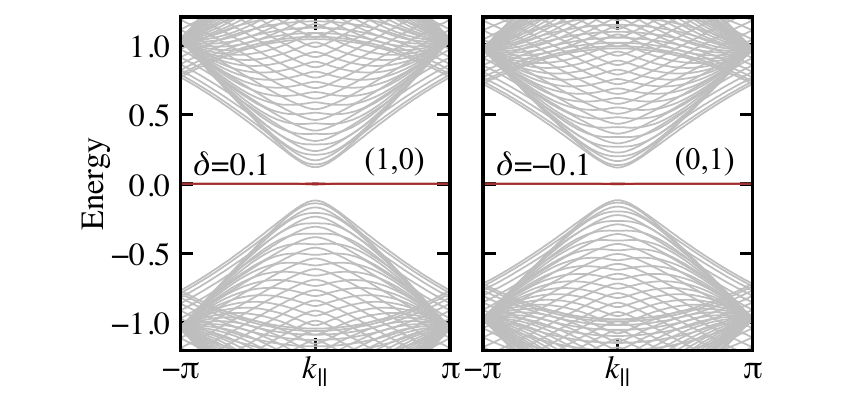}
  \caption{(Color online) Same as Fig.~\ref{fig2_molzig+beard}(a) except for that
  \textit{graphene-zigzag} edge is adopted. The flat
  edge modes are highlighted by the brown color.}\label{fig3_graphenezigzag}
 \end{center}
\end{figure}

\medskip
\noindent\textit{Band structure in ribbon geometry---}
 Now let us evaluate the band structures of a ribbon system long and
 periodic in the direction parallel to $\bm{a}_1$ and of open boundaries
 in the perpendicular direction with the edge shapes illustrated in
 Fig.~\ref{fig1_lattice}. As shown in Fig.~\ref{fig2_molzig+beard}, in
 the case of \textit{molecule-zigzag} edge we find two
 doubly-degenerate dispersive modes associated with helical edge states
 localized at the top and bottom edges for $\delta>0$. The existence of
 two dispersive helical edge states at one edge with zero-energy gap at
 $k_\parallel=0$ are in accordance with $(n_+,n_-)=(1,-1)$ at
 $k_\parallel=0$ and $n_{\rm tot}=0$ for any momentum. For $\delta<0$
 there is no edge mode as expected from the absence of mirror winding
 number in this case. For \textit{partially-bearded} edge with
 $\delta<0$, we observe two dispersive modes associated with helical
 edge states at each of the two edges as shown in
 Fig.~\ref{fig2_molzig+beard} in accordance with the mirror winding
 number (see Table I). In contrary to \textit{molecule-zigzag} edge, in
 the case of \textit{partially-bearded} edge the helical edge states at
 the top and bottom edges are not degenerate due to the absence of
 symmetry with respect to the middle line of ribbon. In contrast to the
 above results, for \textit{graphene-zigzag} edge there is a
 dispersionless flat edge mode for both $\delta>0$ and $<0$ as shown in
 Fig.~\ref{fig3_graphenezigzag}, known for long time in zigzag-edged
 graphene without hopping texture \cite{JPSJ.65.1920}, which is in accordance with $n_{\rm tot}=1$ for any momentum $k_\parallel$ (see
Table I).

  \begin{figure}[tbp]
   \begin{center}
    \includegraphics[width=8.6cm]{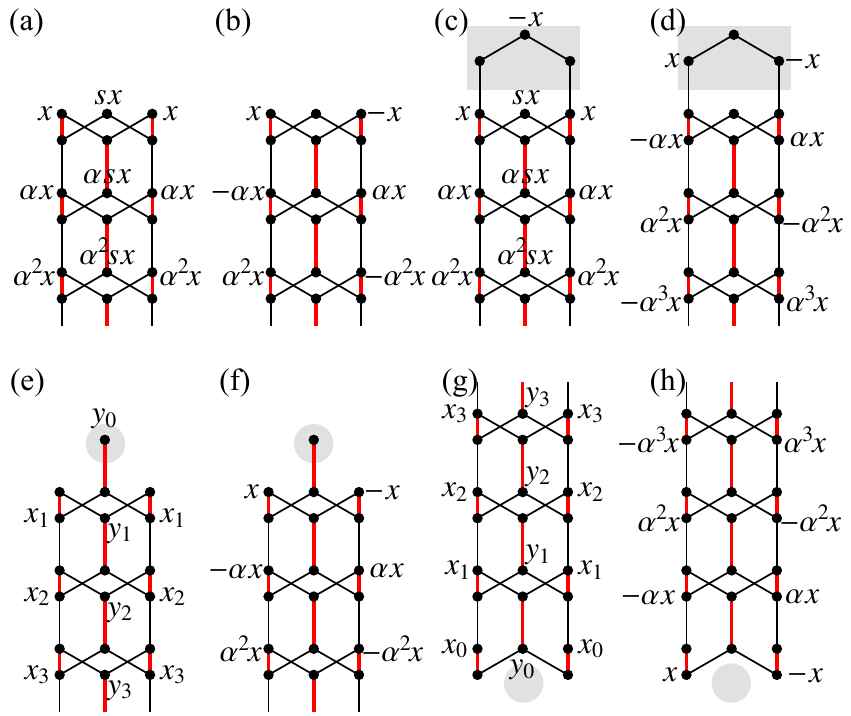}
    \caption{(Color online) Schematic pictures of the
    ansatz for the effective 1D model at $k_\parallel=0$: (a,b) for
    \textit{graphene-zigzag} edge, (c,d) for \textit{molecule-zigzag}
    edge, (e,f)/(g,h) for the top/bottom
    edge of \textit{partially-bearded} edge. (a,c,e,g)/(b,d,f,h) are for
    the even/odd solutions with respect to mirror operation. The
    shaded regions for \textit{graphene-zigzag} and
    \textit{partially-bearded} edges can be regarded as decorations to
    \textit{graphene-zigzag} edge.}
    \label{fig4_edgesolution}
   \end{center}
  \end{figure}

  \medskip
  \noindent\textit{Analytic wave functions of zero-energy edge
  modes---}
The effect of edge decoration becomes more obvious when we look into the
  detailed wave functions of zero-energy edge modes. Taking advantage of
  the chiral symmetry, the wave functions of zero-energy edge modes can
  be derived analytically based on the ansatz $\Psi_A=(\psi_A,0)$ or
  $\Psi_B=(0,\psi_B)$, which have finite weights only on either A or B
  sublattice. Since we focus on $k_\parallel=0$, the system is reduced
  to an effective 1D model as illustrated in
  Fig.~\ref{fig4_edgesolution}. The key point is that, because of the
  mirror
  symmetry, the solutions can be classified by the parity with respect
  to the mirror operation, which puts constraints on and help us find
  the ansatz solutions (see Fig.~\ref{fig4_edgesolution}). 

  Let us begin with the even-parity solution for
  \textit{graphene-zigzag} edge as given in
  Fig.~\ref{fig4_edgesolution}(a), which is symmetric with respect to
  the central column. Since the second site on the left (or right)
  column is connected to three sites, in order to have a nontrivial
  zero-energy mode with $x\neq 0$, one needs to require
  $(s+\alpha)t_0+t_1=0$ \cite{PhysRevB.54.17954,PhysRevB.81.033403}. 
In the same way, from the fourth site on the central column, one has
  $2t_0+\alpha{s}t_1=0$. These two equations are satisfied by $\alpha =
  -\beta+\sqrt{\beta^2+1/\beta}$ and $s = -\beta-\sqrt{\beta^2+1/\beta}$
  with $\beta\equiv(1+\delta)/2$. The solution is physically meaningful
  in the present case only when it decays from the edge into the bulk,
  namely $\alpha<1$, which is achieved by $\delta>0$. For the odd-parity
  solution given in Fig.~\ref{fig4_edgesolution}(b), the ansatz gives a
  solution with $\alpha= 1+\delta$ by inspection noting that the wave
  function is zero at the central column due to symmetry. A meaningful
  edge state therefore exists for $\delta<0$. These results are in
  accordance with the winding numbers summarized in Table I.

  Next we check \textit{molecule-zigzag} edge. As can be seen in
  Fig.~\ref{fig4_edgesolution}(c), the even-parity solution in this case
  is essentially the same as that given in
  Fig.~\ref{fig4_edgesolution}(a) for \textit{graphene-zigzag} edge, for
  which one has a physical solution for $\delta>0$. On the other hand,
  for the odd-parity solution Fig.~\ref{fig4_edgesolution}(d), the
  ansatz gives a solution with $\alpha=1/(1+\delta)$, which is physical
  for $\delta>0$. That is, the even- and odd-parity solutions are both
  physical when and only when $\delta>0$, matching the mirror winding
  number in Table I.

  The \textit{partially-bearded} edge is different from the above two
  cases in the sense the top and bottom edges in ribbon geometry is
  asymmetric with respect to the middle line of the ribbon [see
  Fig.~\ref{fig4_edgesolution}(e) vs. (g) and (f) vs. (h)]. The
  odd-parity solutions in Figs.~\ref{fig4_edgesolution}(f) and
  \ref{fig4_edgesolution}(h) are the same as the one for
  \textit{graphene-zigzag} edge in Fig.~\ref{fig4_edgesolution}(b),
  which are physical for $\delta<0$. For the even-parity solutions as
  given in Figs.~\ref{fig4_edgesolution}(e) and (g), one expects
  that they are physical for $\delta<0$ judging from the mirror winding
  number in Table I, which can be proven with some algebra (see
  appendix). 

  One can derive \textit{molecule-zigzag} edge by adding three sites as
  shown by the shade parts in Figs.~\ref{fig4_edgesolution}(c) and (d)
  to \textit{graphene-zigzag} edge, whereas \textit{partially-bearded}
  edges are obtained by adding or deleting one site [see the shade parts
  in Figs.~\ref{fig4_edgesolution}(e-h)]. It is then interesting to
  observe that the decoration of \textit{graphene-zigzag} edge leading
  to \textit{molecule-zigzag} (\textit{partially-bearded}) edge only
  affects the odd-parity (even-parity) solution. For
  \textit{graphene-zigzag} edge, one has one zero-energy mode for both
  $\delta>0$ and $\delta<0$. The decoration to \textit{graphene-zigzag}
  edge then yields two zero-energy modes for $\delta>0$ ($\delta<0$) for
  \textit{molecule-zigzag} (\textit{partially-bearded}) edge, and
  suppresses the zero-energy mode for $\delta<0$ ($\delta>0$). The
  coexisting zero-energy even- and odd-parity solutions at $\Gamma$
  point are mixed for $k_\parallel\neq 0$, where the mirror symmetry is
  no longer effective. Because of the
  parity difference, the coupling is linear in $k_\parallel$ and
  generates two linearly dispersive modes, which are nothing but the
  helical edge modes shown in
  Figs.~\ref{fig2_molzig+beard}.

  \medskip
  \noindent\textit{Discussions ---}
  In the previous work\cite{Wu:2016aa}, the parameter regime $t_1<t_0$
  was regarded as topologically trivial. However, as summarized in Table
  I this parameter regime supports a state characterized by nonzero
  mirror winding number. The difference comes from the different choices
  of unit cell: in the previous work the circular unit cell was taken,
  while the rhombic one is revealed to be appropriate (see
  Fig.~\ref{fig1_lattice}). The way of taking unit cell is more than a
  convention in the present approach for realizing topological states
  based on hopping texture, since the existence of topological edge
  modes, and thus whether the system is topological or trivial,
  crucially relies on the shape of edge which is uniquely related to the
  unit cell. In this sense, the present work illuminates a way to
  realize topological states by edge decoration, which is expected very
  useful for tailoring graphene.
  
To summarize, topological phases in honeycomb lattice induced by texture
in hopping energy between nearest-neighboring sites are characterized in
terms of the mirror winding number, a rigorous topological invariant
intimately related to the chiral (or sublattice)
symmetry. Analytic wave functions are provided for zero-energy
edge modes at the $\Gamma$ point, which evolve into the
dispersive helical edge states at finite momenta. Explicitly,
when the intra-hexagon hopping energy is stronger (weaker) than the
inter-hexagon one, the \textit{molecule-zigzag}
(\textit{partially-bearded}) edge to honeycomb lattice yields gapless
helical edge modes. The present work provides a new designing guideline
for topological edge states adaptive to the bulk hopping texture, which
may pave a way to tailoring graphene in the topological point of view.

\medskip
\begin{acknowledgments}
 \noindent\textit{Acknowledgments---}
 XH is grateful for the stimulating discussions with D.~Haldane, E.~Lieb
 and L.~Fu. This work was supported by the WPI Initiative on Materials
 Nanoarchitectonics, Ministry of Education, Cultures, Sports, Science and
 Technology, Japan. 
\end{acknowledgments}

\appendix

\section{Appendix}

 \subsection{Armchair Edge}
\begin{figure}[h]
   \begin{center}
    \includegraphics[scale=1.0]{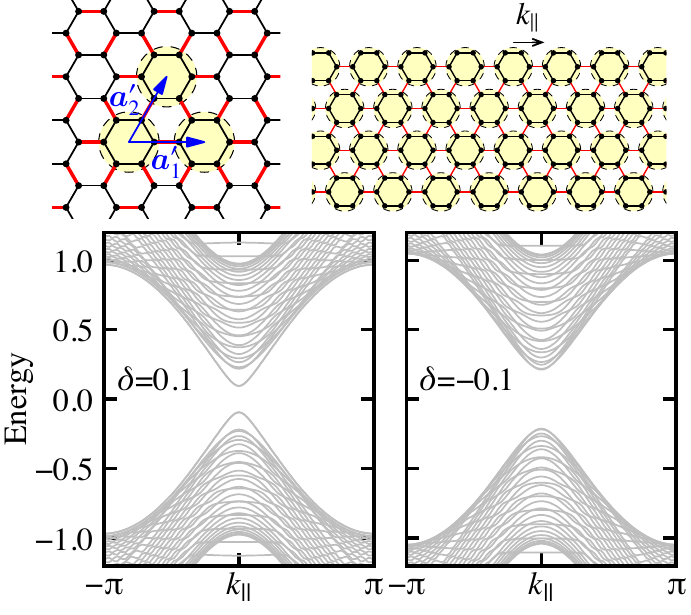}
    \caption{Same as Fig.~2 in the main text except for that armchair
    edge is adopted, and that the width is 20 unit cells along
    $\bm{a}'_2$.}\label{fig:armchair}
   \end{center}
 \end{figure}

 Mirror symmetry also exists for the armchair edge, a case well studied
 for graphene. However, considering the way to take unit cell compatible
 with armchair edge as displayed in Fig.~\ref{fig:armchair}, one finds
 that the mirror operation perpendicular to armchair edge does not
 commute with the chiral operator
 $\gamma$ defined in the main text, since the mirror operation in this
 case mixes A and B sublattices. Therefore, no mirror winding number can
 be assigned in this case, and as the result, no zero-energy edge modes
 exist for a ribbon with armchair edge, which agrees with the
band calculation as shown in Fig.~\ref{fig:armchair}.
 
 \subsection{Evaluation of the Winding Number}
 Here we provide some details for the evaluation of winding numbers.
 Let us consider an effective 1D model constructed from a
 2D model such as that given in the main text [see Eq.~(2) in main
 text]
 \begin{equation}
  H_{k_\parallel,k_\perp}
   =
   \begin{pmatrix}
    0&Q_{k_\parallel,k_\perp}\\
    Q^\dagger_{k_\parallel,k_\perp}&0
   \end{pmatrix}
 \end{equation}
 by regarding $k_\parallel$ as a free parameter. Then, 
 the winding number $w$ is evaluated as \cite{doi:10.7566/JPSJ.85.022001}
\begin{align}
 w&=\frac{1}{4\pi{i}}\int_0^{2\pi}\mathrm{tr}
 \Bigl[\gamma{H}_k^{-1}\frac{dH_k}{dk}\Bigr]dk\nonumber\\
 &=-\frac{1}{2\pi}\int_0^{2\pi}\frac{d}{dk}\arg(\det{Q}_k)dk
 \label{formula:winding}
\end{align}
with $k_\perp$ denoted by $k$ and $k_\parallel$ suppressed for clarity, 
where the second line follows from the relations
\begin{equation}
 H_k^{-1}
  =
  \begin{pmatrix}
   0&(Q^\dagger_k)^{-1}\\
   Q_k^{-1}&0
  \end{pmatrix},
\end{equation}
and $\mathrm{tr}Q^{-1}_kdQ_k=d\log\det{Q}_k$.
Physically, Eq.~\eqref{formula:winding} detects how $\det{Q}_k$
winds about the origin of the complex plane while $k$ evolves from $0$ to
$2\pi$. 

\subsubsection{\textit{Mirror winding number for molecule-zigzag edge}}
The winding number for \textit{molecule-zigzag} edge in the odd sector
$n_-=-1$ for $t_1>t_0$, whereas $n_-=0$ for $t_1<t_0$ can be obtained
straightforwardly from the second equality in Eq.~(4) in the main text
\cite{PhysRevLett.49.1455}. For the even sector, one can see that, for
$t_1=0$,
$n_+=0$ since there is no momentum dependence in $Q^+_{k_\perp}$ in the
first equality of Eq.~(4) in the main text. This topological property
remains unchanged in the whole regime $t_1<t_0$ where the gap always
keeps open. On the other hand, for $t_0=0$, one has $n_+=1$ since one
component of $Q^+_{k_\perp}$ winds twice about the origin of the complex
plane as $k_\perp$ evolves from 0 to $2\pi$, while the other component
winds once in the opposite direction. This winding number also remains
the same for the whole regime $t_1>t_0$.
One can confirm this intuitive discussion by plugging 
\begin{equation}
 \det{Q^+_{k_\perp}}=t_1\bar{Y}^2(t_0+t_1Y)-2t_0^2
\end{equation}
into Eq.~\eqref{formula:winding} and performing integration. 
The resultant winding numbers
are summarized in Table I in the main text.

\subsubsection{\textit{Mirror winding number for partially-bearded edge}}
Next we consider \textit{partially-bearded} edge.  For the odd sector
the analysis is again obvious, noticing that in the second equality of
Eq.~(6) in the main text, $t_0$ and $t_1$ switch their positions from
those in Eq.~(4), which yields $n_-=1$ for $t_1<t_0$, whereas $n_-=0$
for $t_1>t_0$. For the even sector, we have
\begin{equation}
 \det{Q}_{k_\perp}^{+}
  =t_1(t_1+t_0\bar{Y})-2t_0^2Y.
\end{equation}
Then, it is clear that one has $n_+=0$ for $t_1>t_0$ as can
be seen from the limit $t_0=0$, while $n_+=-1$ seen from the limit
$t_1=0$. This of course can be confirmed by directly performing
integration in Eq.~\eqref{formula:winding}. 

 \subsubsection{\textit{Mirror winding number for graphene-zigzag edge}}
 It is straightforward to see that for \textit{graphene-zigzag} edge,
 one has
 \begin{equation}
 Q_{\bm{k}}=
  \begin{pmatrix}
   t_1X\bar{Y}&t_0&t_0\\
   t_0&t_1\bar{X}&t_0\bar{Y}\\
   t_0&t_0\bar{Y}&t_1
  \end{pmatrix},
 \end{equation}
which yields
 \begin{equation}
   Q^+_{k_\perp}=
  \begin{pmatrix}
   t_1\bar{Y}&\sqrt{2}t_0\\
   \sqrt{2}t_0&t_1+t_0\bar{Y}
    \end{pmatrix},\quad
    Q^-_{k_\perp}=t_1-t_0\bar{Y}.
 \end{equation}
 Again, the winding number in the odd sector is obvious. For the even
 sector, using
 \begin{equation}
  \det{Q}^+_{k_\perp}=t_1\bar{Y}(t_1+t_0\bar{Y})-2t_0,
 \end{equation}
 one can obtain $n_+=1$ for $t_1>t_0$ by considering the limit case
 $t_0=0$, and $n_+=0$ for $t_1<t_0$ by taking the limit $t_1=0$.

 \subsection{Analytic Solution for \textit{Partially-Bearded} Edge}
 Here we discuss the wave functions of the even-parity zero-energy modes
 for \textit{partially-bearded} edge. For the top edge as shown in
 Fig.~4(e) in the main text, in order to have a nontrivial solution one
 has to require $2t_0x_{j+1}+t_1y_j=0$ (see the central column) and
 $t_1x_{j+1}+t_0y_{j+1}+t_0x_j=0$ (see the left or right column) for
 $j>0$. They can be summarized into a matrix form
 \begin{equation}
 \begin{pmatrix}
  x_{j+1}\\
  y_{j+1}
 \end{pmatrix}
 =A
 \begin{pmatrix}
  x_j\\
  y_j
   \end{pmatrix},\quad
   A =
    \begin{pmatrix}
  0 & -\beta \\
 -1 & 2\beta^2
 \end{pmatrix},
 \end{equation}
 with $\beta=(1+\delta)/2$ as defined in the main text, which gives
 \begin{equation}
 \begin{pmatrix}
  x_n\\
  y_n
 \end{pmatrix}
 =A^{n-1}
 \begin{pmatrix}
  x_1\\
  y_1
 \end{pmatrix}.
 \end{equation}
 The two eigenvalues of the matrix $A$ are given by
 $\epsilon_\pm=\beta^2\pm\sqrt{\beta^4+\beta}$, which should satisfy
 $|\epsilon_{\pm}|<1$ in order to make the solution decaying into the
 bulk. This is guaranteed by $\delta<0$. The same discussions apply for
 the bottom edge shown in Fig.~4(g) in the main text.

\end{document}